\documentclass{lhep}
\usepackage{graphicx}
\usepackage{cite}

\journal{LHEP}
\vol{xx}
\jyear{2018}
\pages{xxx}

\received{xx January 2018}
\published{xx March 2018}
\begin{document}
\title{Dimuon enhancement at 28 GeV and tentative \\
(pseudo)scalar partner of the $Z$ boson at 57.5 GeV}
\author{
Eef~van~Beveren\auno{1}
and
George~Rupp\auno{2}
}
\address{$^1$Centro de F\'{\i}sica da UC,\\
Departamento de F\'{\i}sica,
Universidade de Coimbra, P-3004-516 Coimbra, Portugal}
\address{$^2$Centro de F\'{\i}sica e Engenharia de Materiais Avan\c{c}ados,\\
Instituto Superior T\'{e}cnico,
Universidade de Lisboa, P-1049-001 Lisboa, Portugal}
\begin{abstract}
The CMS Collaboration at the LHC recently reported
an accumulation of data around 28 GeV in the invariant-mass
distribution of muon pairs in association with a $b$ quark jet and
at least a second jet.
This is analysed here in the light of the possible existence of
a (pseudo)scalar boson with mass of about 57.5 GeV.
We find that part of the data may originate
in the radiative decay of $Z$ bosons
into pairs consisting of the lighter boson and a photon, giving rise
to dimuon decay products that either stem from the photon or from the
(pseudo)scalar boson.
\end{abstract}
\maketitle
\begin{keyword}
PACS 13.38.Dg, 13.30.Ce, 13.66.Lm, 12.90.+b
\end{keyword}

In three previous papers
\cite{ARXIV13047711,EPJWC95p02007,APPS8p145}
we argued that data indicate the possible existence
of a (pseudo)scalar boson with mass of about 57.5 GeV.
In Ref.~\cite{ARXIV13047711} we focused on a dip at about 115 GeV
in diphoton data published \cite{PLB710p403} by the Compact-Muon-Solenoid
(CMS) Collaboration at the Large Hadron Collider (LHC),
corroborated by similar data by the ATLAS Collaboration
\cite{PLB716p1}. We interpreted this dip as the threshold
for the production of a pair of (pseudo)scalar bosons
with a mass of about 57.5 GeV.
In the following we will refer to such a boson as ${Z_0}$.

\begin{figure}[htbp]
\begin{center}
\begin{tabular}{c}
\includegraphics[height=120pt]{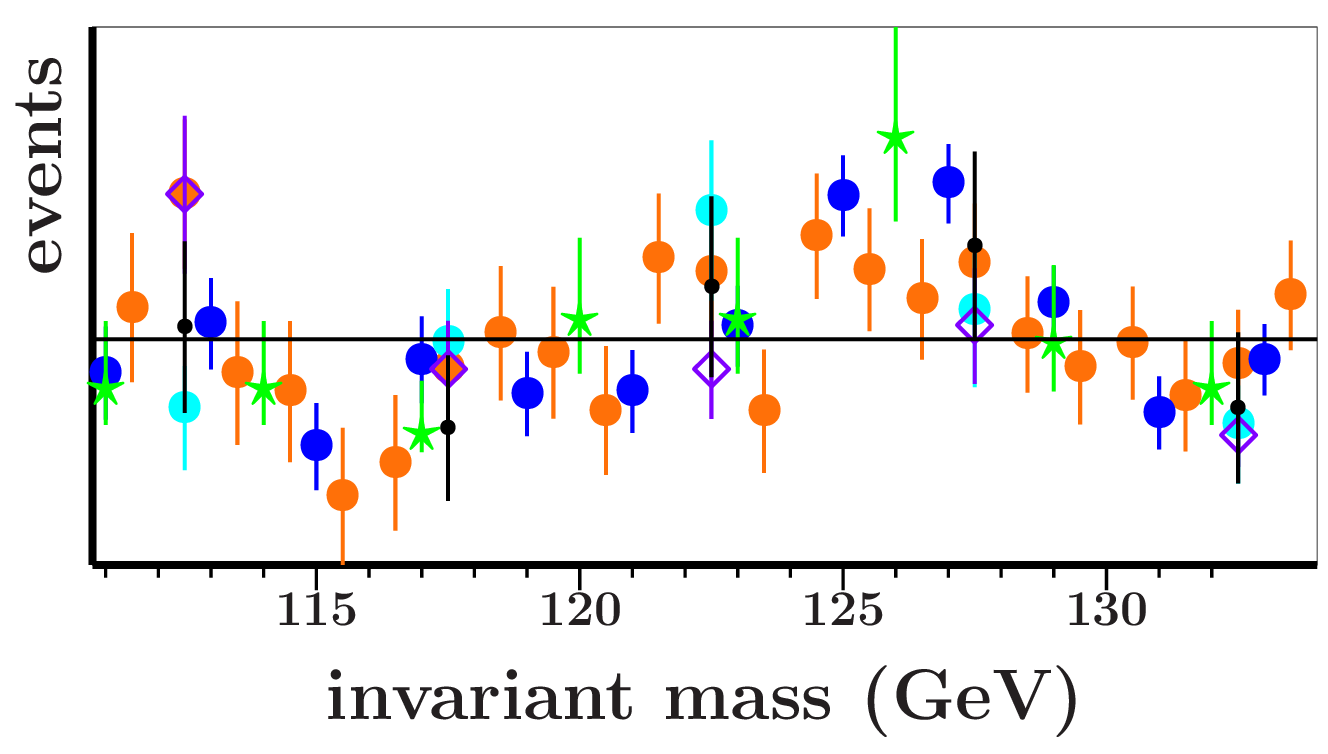}
\end{tabular}
\end{center}
\caption[]{Diphoton signals published
by CMS \cite{CMSPASHIG-13-016}
({\definecolor{tmpclr}{rgb}{1.000,0.440,0.030}{\color{tmpclr}$\bullet$}})
and ATLAS
\cite{ARXIV13053315} ({\color{blue}$\bullet$}),
four-lepton signals
by CMS Collaboration \cite{PRD89p092007}
({\color{green}$\star$})
and ATLAS \cite{ARXIV13053315}
({\definecolor{tmpclr}{rgb}{0.000,1.000,1.000}{\color{tmpclr}$\bullet$}}),
invariant-mass distributions for
$\tau\tau$ in $e^{+}e^{-}\to\tau\tau (\gamma )$
({\definecolor{tmpclr}{rgb}{0.500,0.000,1.000}{\color{tmpclr}$\diamond$}})
and
$\mu\mu$ in $e^{+}e^{-}\to\mu\mu (\gamma )$ ({$\bullet$})
by L3 \cite{PLB479p101}.
}
\label{dip115}
\end{figure}

In Ref.~\cite{EPJWC95p02007} we showed that the dip at about 115~GeV
in the diphoton data of the CMS and ATLAS Collaborations
is also corroborated by four-lepton signals published
by CMS \cite{PRD89p092007} and ATLAS \cite{ARXIV13053315}.
This dip is compatible with older
data for $\tau\tau$ in $e^{+}e^{-}\to\tau\tau (\gamma )$
and $\mu\mu$ in $e^{+}e^{-}\to\mu\mu (\gamma )$
by the L3 Collaboration \cite{PLB479p101}.
All data, shown in Fig.~\ref{dip115},
indeed seem to indicate that the total signal
in the 115--133 GeV mass interval is in fact the sum of
two different contributions
viz.\ a broad non-resonant threshold enhancement
preceded by a sharp dip at about 115 GeV
and a resonance around 125 GeV.

The idea that the heavy gauge bosons may be composite
and so must have spin-zero partners as well,
the latter with a mass in the range 40--60~GeV,
was considered long ago \cite{PLB135p313}
and studied more generally in many papers (see e.g.\ Refs.\
\cite{PLB141p455,PRAMANA23p607,PRD36p969,NCA90p49,PRD39p3458,PRL57p3245,
IJMPA30p1550014,ARXIV160705403,MPLA32p1750057}).
So far no experiment has produced evidence of their existence.
However, recently the interest in weak substructure has been renewed
\cite{ARXIV12074387,ARXIV12105462,ARXIV13076400,ARXIV13040255,PRD90p035012},
also due to the
observation of the Higgs boson at the LHC \cite{PLB716p1,PLB716p30}.

In Ref.~\cite{MPLA32p1750057} a comparison
of gauge-boson partner states to mesons with matching quantum numbers
led to mass predictions of the order of 300 GeV,
with the exception of the scalar state,
being indentified with the Higgs.
The corresponding pseudoscalar boson,
not discussed in Ref.~\cite{MPLA32p1750057},
would then most likely be lighter than the $W^{\pm}$ and $Z$,
just like the pion is (much) lighter than the $\rho$ meson.
Furthermore, in the meson sector the scalar $f_{0}(500)$
(alias $\sigma$) is even lighter than the $\rho$.

\begin{figure}[hbtp]
\centering
\begin{tabular}{c}
\includegraphics[width=220pt]{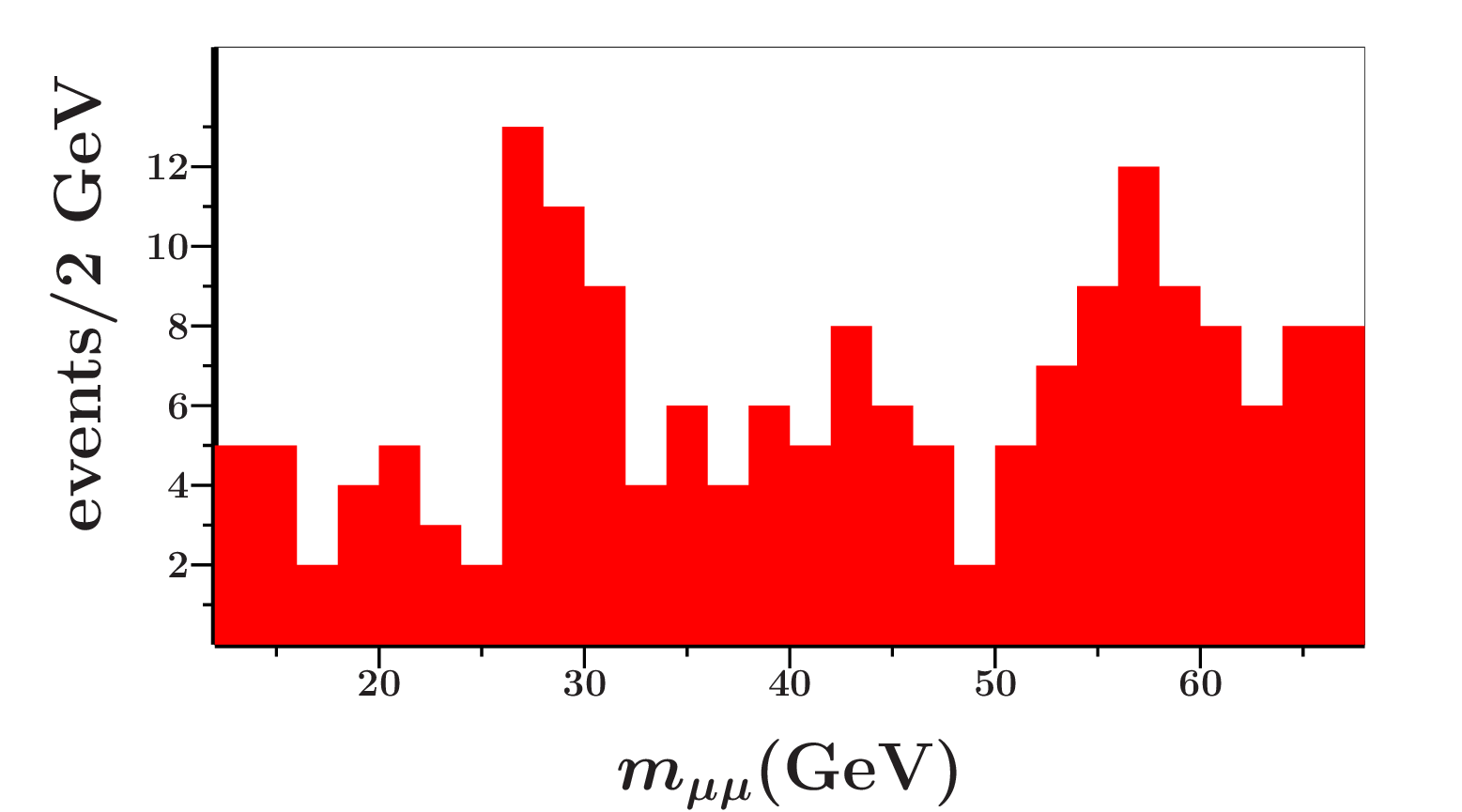}
\end{tabular}
\caption[]{\small
Data on the dimuon mass distribution in $Z$ decays,
taken from Ref.~\cite{ARXIV180801890}.
}
\label{cms}
\end{figure}

Now, very recently the CMS Collaboration reported \cite{ARXIV180801890}
on a data analysis that appears to support the hypothesis of a
$Z\to\gamma Z_{0}$ decay process, with a $Z_{0}$ mass of 57.5 GeV
and a photon energy of about 28 GeV.
Namely, in the latter study
an excess of events above the background was observed
near a dimuon invariant mass of 28 GeV,
with a significance of 4.2 standard deviations.
The data had been collected in 2012 with the CMS detector
in proton-proton collisions at the LHC,
for centre-of-mass (CM) energies of 8 TeV and
with an integrated luminosity of 19.7 fb$^{-1}$.
The event selection required a $b$-quark jet, with at least one jet
in the central and the forward pseudorapidity region, respectively.
The result is depicted in Fig.~\ref{cms}.

The reader should be aware that the data in Fig.~\ref{cms}
are to be considered with some caution.
In a different dimuon data selection with higher statistics,
the CMS Collaboration obtained a signal near 28 GeV
with a significance of only 2.9 standard deviations.
Moreover, in related dimuon events
selected from data collected at the LHC in 2016,
for proton-proton collisions at CM energies of 13 TeV
and corresponding to an integrated luminosity of 35.9 fb$^{-1}$,
CMS found near 28 GeV signals of 2.0 standard deviations and a 1.4
standard-deviation deficit for the two mutually exclusive
dimuon-event categories.
Accordingly, CMS concluded \cite{ARXIV180801890}
that more data and additional theoretical input are required
to understand the results.

At present there are in the literature interpretations that the CMS dimuon
signal may explain the discrepancy between the experimentally measured
and theoretically predicted muon anomalous magnetic moment
\cite{ARXIV180802431}, or might be modelled as
a pseudoscalar mixture of toponium and gluonium \cite{ARXIV160100618}.

A closer look at the data depicted in Fig.~\ref{cms}
reveals a second accumulation of data near 57 GeV.
Now, if we assume that the $Z_{0}$ does exist,
then in the reaction $Z\to\gamma Z_{0}$ the $Z_{0}$ and the intermediate
photon have masses of 57.5 and 28 GeV, respectively.
Moreover, both particles couple to dimuons.
This would partly explain why the data show two enhancements,
i.e., one near 28 GeV and one near 57 GeV.
We say ``partly'', because one would expect the
$Z_{0}$ to be much less likely to decay into muon pairs than the photon.
Actually, one expects the $Z_{0}$ to dominantly couple to $\gamma\gamma$.

\begin{figure}[htbp]
\begin{center}
\begin{tabular}{cc}
\includegraphics[height=100pt]{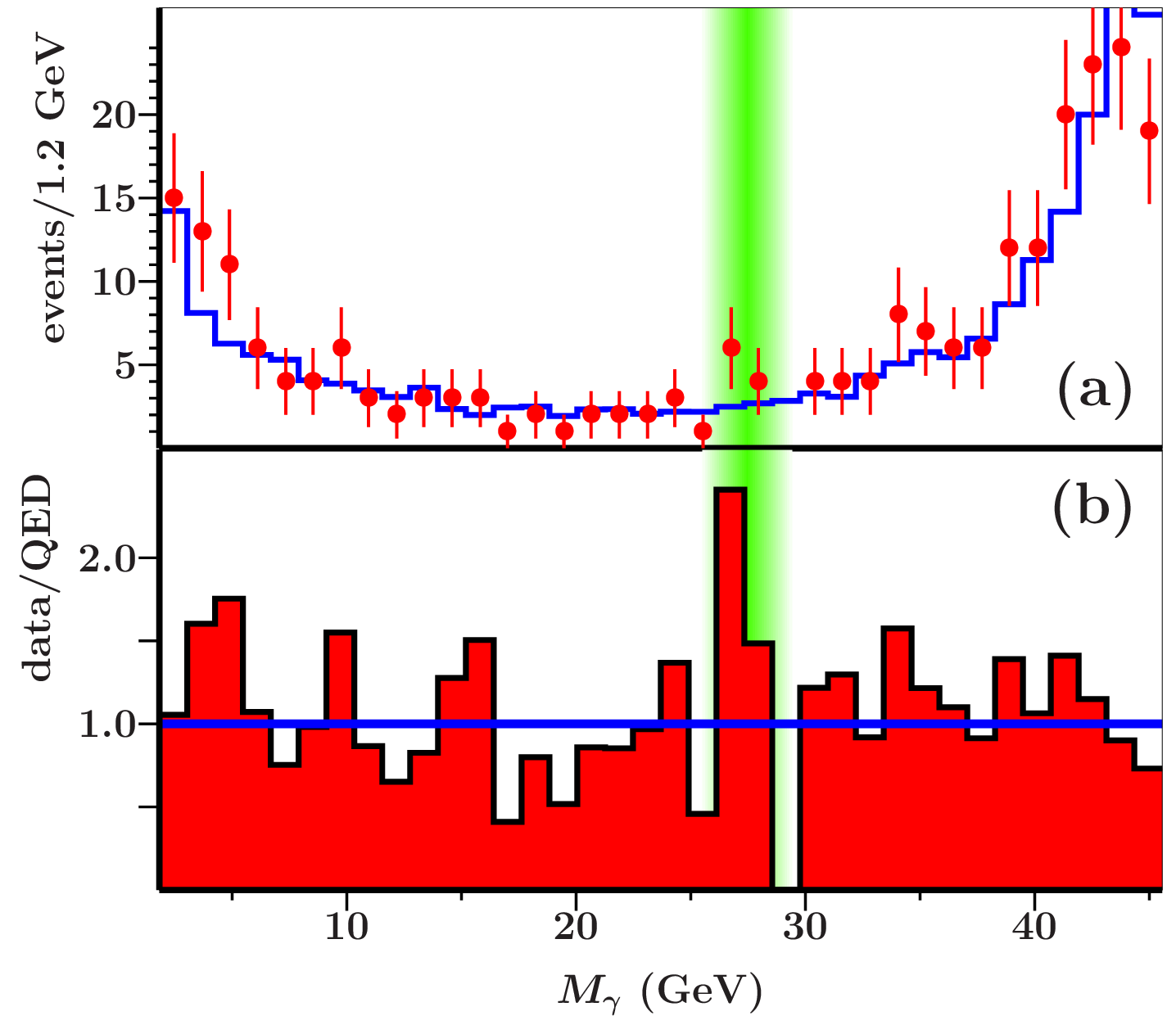} &
\includegraphics[height=100pt]{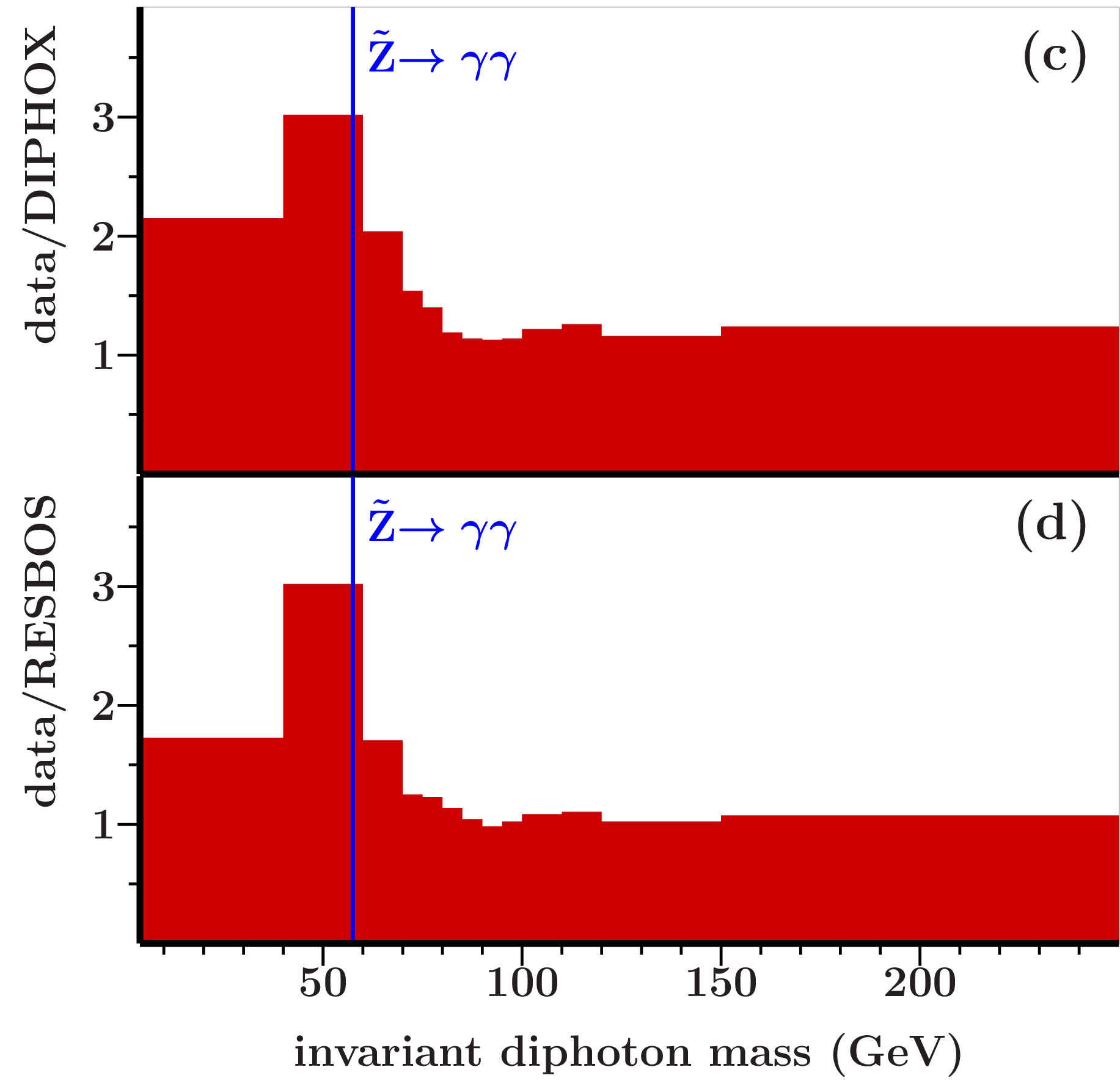}
\end{tabular}
\end{center}
\caption[]{
(a): Experimental data for the three one-photon CM energies
of the 87 candidate $Z\to 3\gamma$ events
measured by the L3 Collaboration \cite{PLB345p609},
assuming $\sqrt{s}=M_{Z}$.
The histogram was obtained by L3
from a Monte-Carlo simulation for the expected number of events
predicted by QED.
With the green band we indicate where we expect
photons from the radiative process $Z\to\gamma Z_{0}$
for the case that $Z_{0}$ has a mass of 57.5 GeV.
(b): The same data as shown in (a), but now measured events
divided by QED-expected events.
(c, d): Measured over expected events for diphoton invariant-mass
distributions published by the CMS Collaboration \cite{CMSPASHIG-13-001},
for (c) DIPHOX and (d) RESBOS.
}
\label{1photonL3}
\end{figure}

In Fig.~\ref{1photonL3} we collect data
for the reaction $Z\to\gamma Z_{0}\to\gamma\gamma\gamma$
published by the L3 Collaboration \cite{PLB345p609}
and for the diphoton mass distribution from the CMS Collaboration
\cite{CMSPASHIG-13-001}.
The L3 Collaboration used 65.8 pb$^{-1}$ of data
taken during the 1991--1993 runs at the Large Electron-Positron Collider
(LEP) on top of and around the $Z$ peak, for CM energies
between 88.5 and 93.7 GeV.
On the other hand, the CMS analysis is based on 5 fb$^{-1}$
isolated diphoton production cross sections
collected at 7 TeV $pp$ CM energies in the year 2011.

In Fig.~\ref{1photonL3}a/b we depict the L3 data for
the three one-photon CM energies for each of the candidate events.
L3 expressed the one-photon CM energies
as a function of $M_{\gamma}/\sqrt{s}$.
Here, we convert this information into $M_{\gamma}$,
while assuming $\sqrt{s}=M_{Z}$.
Moreover, we indicate where we expect the photons from the radiative
process $Z\to\gamma Z_{0}$ by a green band.
One sees that most of the L3 data agree well with the expectation from QED.
Nevertheless, be it a coincidence or not, in the mass region
where we expect a signal from $Z\to\gamma Z_{0}$ events,
we observe a small enhancement.
This can be demonstrated better by showing
the ratio of measured signal over QED prediction,
as depicted in Fig.~\ref{1photonL3}b.
Now one clearly observes a modest enhancement,
for exactly the expected $Z_{0}$ mass of 57.5 GeV.

Finally, in Fig.~\ref{1photonL3}c we show
diphoton invariant-mass distributions
measured by the CMS Collaboration \cite{CMSPASHIG-13-001}
for predictions of a DIPHOX data simulator and
in Fig.~\ref{1photonL3}d for CMS predictions using RESBOS.
In both figures one again observes an excess of three times
more data than predicted in the 40--60 GeV mass interval.
This could be in agreement with diphotons
stemming from the reaction $Z_{0}\to\gamma\gamma$.

Summarising, we have shown in the foregoing
that our prior work on a possible weak substructure
\cite{ARXIV13047711,EPJWC95p02007,APPS8p145}
is compatible with the observed enhancements
in dimuon invariant-mass distributions
recently reported by the CMS Collaboration \cite{ARXIV180801890}.
Moreover, the decay mode $Z_{0}\to\mu\mu$ favours
a pseudoscalar assignment for the $Z_{0}$.

\newcommand{\pubprt}[4]{#1 {\bf #2}, #3 (#4)}
\newcommand{\ertbid}[4]{[Erratum-ibid.~#1 {\bf #2}, #3 (#4)]}
\def\APPS{Acta Phys.\ Polon.\ Supp.}
\def\EPJWC{Eur.\ Phys.\ J.\ Web of Conf.}
\def\IJMPA{Int.\ J.\ Mod.\ Phys.\ A}
\def\MPLA{Mod.\ Phys.\ Lett.\ A}
\def\NCA{Nuovo Cim.\ A}
\def\PLB{Phys.\ Lett.\ B}
\def\PRAMANA{Pramana J.\ Phys.}
\def\PRD{Phys.\ Rev.\ D}
\def\PRL{Phys.\ Rev.\ Lett.}

\end{document}